\documentclass{article}
\usepackage{amsmath}
\usepackage{amsfonts}
\usepackage{graphicx}
\usepackage{chemarrow}
\usepackage{setspace}
\begin{document}

\title{A Much better replacement of the Michaelis-Menten equation and its application}
\author{$Banghe\ Li^{a,b,*}$
\and $Bo\ Li^{a,c}$ \and $Yuefeng\ Shen^{a,d}$}
\maketitle
\begin{abstract}
Michaelis-Menten equation is a basic equation of enzyme kinetics and
gives an acceptable approximation of real chemical reaction
processes. Analyzing the derivation of this equation yields the fact
that its good performance of approximating real reaction processes
is due to Michaelis-Menten curve (\ref{QSSL2equal0}). This curve is
derived from Quasi-Steady-State Assumption(QSSA), which has been
proved always true and called Quasi-Steady-State Law by Banghe Li et
al [\ref{refBanghe2008}].
\par
Here, we found a quartic equation $A(S,E)=0$
(\ref{replacementofMM}), which gives more accurate approximation of
the reaction process in two aspects: during the quasi-steady state
of a reaction, Michaelis-Menten curve approximates the reaction
well, while our quartic equation $A(S,E)=0$ gives better
approximation; near the end of the reaction, our equation approaches
the end of the reaction with a tangent line same to that of the
reaction, while Michaelis-Menten curve does not. In addition, our
quartic equation $A(S,E)=0$ differs to Michaelis-Menten curve less
than the order of $1/S^3$ as $S$ approaches $+\infty$.
\par
By considering the above merits of $A(S,E)=0$, we suggest it as a
replacement of Michaelis-Menten curve. Intuitively, this new
equation is more complex and harder to understand. But, just because
its complexity, it provides more information about the rate
constants than Michaelis-Menten curve does.
\par
Finally, we get a better replacement of the Michaelis-Menten
equation by combing $A(S,E)=0$ and the equation $dP/dt=k_2C(t)$.

\end{abstract}
$keywords:$ rate constants of enzyme kinetics; quasi-steady-state assumption; quasi-steady-state law.

\vspace{20pt}

a. Key Laboratory of Mathematics
Mechanization, Academy of Mathematics and Systems Science, Chinese
Academy of Sciences, Beijing 100190.

b. libh@amss.ac.cn. Family name: Li. Telephone number:
086-010-62651273.

c. libo@amss.ac.cn. Family name: Li.

d. shyf@amss.ac.cn. Family name: Shen.

* Corresponding author

\newpage

\doublespacing

\section{Introduction}

Enzymes are biological catalysts in almost all life processes.
Enzyme kinetics as an important branch of enzymology studies the
rate of reaction and the change of rate under different conditions.
It is essential to describe the reaction
mechanism[\ref{refVoet1999}].
\par
In 1902, Adrian Brown studied the rate of hydrolysis of sucrose by
yeast enzyme $\beta$-fructofuranosidase, which was considered as the
first case study of enzyme kinetics[\ref{refBrown1902}]. Victor
Henri proposed two reaction mechanisms which contains only one
substrate and one product forming a substrate-enzyme
complex[\ref{refHenri1902}, \ref{refSchnell2006}]. One of them
became the basic model of enzyme kinetics:

\begin{equation}\label{ChemicalReactionScheme }
E+S \autorightleftharpoons{$k_1$}{$k_{-1}$} C
\autorightarrow{$k_2$}{} P + E,
\end{equation}
where $E$, $S$, $C$, $P$ represent enzyme, substrate,
enzyme-substrate complex and product, respectively. And $k_1$,
$k_{-1}$, $k_2$ represent the rate constants of corresponding
reaction steps.
\par
Since Briggs and Haldane proposed the quasi-steady-state-assumption
(QSSA) in 1925[\ref{refBriggsHaldane1925}], this simplest model has
been thoroughly studied under QSSA[\ref{refVoet1999},
\ref{refFersht1985}, \ref{refSchulz1994}]. By QSSA, Briggs and
Haldane obtained the classic Michaelis-Menten equation:
\begin{equation*}\label{MichaelisMentenEquation}
    v_0=V_{\max}S_0/(K_{M}+S_0),
\end{equation*}
where $v_0$ is the initial velocity of the reaction, $K_M$ is the
Michaelis constant defined as $K_M=(k_{-1}+k_2)/k_1$ and $V_{\max}$
is the so-called maximal velocity in many literatures, which is
actually the supremum of the velocity but is never reached.
Michaelis-Menten equation soon became the basic equation of enzyme
kinetics[\ref{refVoet1999}]. All the experimental results so far
show that Michaelis-Menten equation provides a good description of
enzyme kinetics processes for large ensemble of enzyme molecules
when the concentration of substrate exceeds that of enzyme greatly.
At the single-molecule level, the enzyme molecule moves according to
thermal fluctuation and reacts stochastically with substrate
molecules[\ref{refXie1999}, \ref{refQian2002}]. By statistical
analysis of the stochastic behaves, Michaelis-Menten equation also
holds[\ref{refAranyi1977}, \ref{refXie2006}].
\par
After Briggs and Haldane's work, Lineweaver and
Burk[\ref{refLieweaver1934}] found that the reciprocal form of
Michaelis-Menten equation gave a linear relation between $1/v_0$ and
$1/S_0$, i. e.

\begin{equation}\label{reciprocalMMequation}
   1/v_0=\big(K_M/V_{\max}\big)1/S_0+1/V_{\max}.
\end{equation}
This linear relation can be used to estimate the kinetics parameters
with least square method. Although this estimation sometimes may
lead to relative poor accuracy[\ref{refDowdRiggs1965},
\ref{refChan1995}, \ref{refRitchiePrvan1996}], many textbooks
recognized its value on simplicity and
visualization[\ref{refVoet1999}, \ref{refSegel1975},
\ref{refDixonWebb1979}]. Michaelis-Menten equation do waste too much
information on progress curve. In fact, Michaelis-Menten equation is
derived from the quadratic equation $dE/dt=0$ which can describe the
whole process of the chemical reaction except the initial transient
period provided $S_0\gg E_0$[\ref{refSchnellMendoza1997}].

\par
The validity of the Michaelis-Menten equation is strongly dependent
on the validity of QSSA. Many biologists tested QSSA through
biological experiments or computational experiments. But no one can
confirm its validity during the next 80 years until recently Banghe
Li et al. gave the rigorous description of this assumption and
proved it mathematically[\ref{refBanghe2008}]. Thus, from now on,
this assumption is called the Quasi-Steady-State Law(QSSL).
Moreover, this assures the validity of the Michaelis-Menten
equation. This may be the first application of qualitative theory of
dynamical systems into this basic enzyme kinetics model.

To quote the QSSL, we first introduce the basic model of enzyme
kinetics. The enzyme kinetics is a branch of chemical
kinetics[\ref{refVoet1999}]. Thus, according to the law of mass
action the time evolution of concentrations of reactants is
determined by the following differential
equations[\ref{refSegelSlemrod1989}]:
\begin{eqnarray}
  dS/dt(t) &=& -k_1S(t)E(t)+k_{-1}C(t) \label{eqnds}\\
  dE/dt(t) &=& -k_1S(t)E(t)+(k_{-1}+k_2)C(t) \label{eqnde}\\
  dC/dt(t) &=& k_1S(t)E(t)-(k_{-1}+k_2)C(t) \label{eqndc}\\
  dP/dt(t) &=& k_2C(t) \label{eqndp}
\end{eqnarray}
with the initial condition
\begin{equation}\label{eqninitialcondtion}
    (S(0),E(0),C(0),P(0))=(S_0,E_0,0, 0).
\end{equation}
where $E(t)$, $S(t)$, $C(t)$ and $P(t)$ denote the concentrations of
enzyme, substrate, enzyme-substrate complex and product at time $t$
during the process, respectively. Under the two conservation laws
\begin{eqnarray}
  E(t)+C(t) &=& E_0 \label{conservationlawE}\\
  S(t)+C(t)+P(t) &=& S_0,\label{conservationlawS}
\end{eqnarray}
these differential equations are equivalent to system of
differential equations consisted of $(S(t), E(t))$, $(S(t), P(t))$
or $(P(t), E(t))$, i. e.
\begin{equation}\label{eqnardSdtdEdt}
\left \{
\begin{array}{lll}
dS/dt(t) &=& -k_1S(t)E(t)+k_{-1}(E_0-E(t)) \, \vspace{1.5mm} \\
dE/dt(t) &=& -k_1S(t)E(t)+(k_{-1}+k_2)(E_0-E(t)) \,
\end{array}
\right.,
\end{equation}
\begin{equation}\label{eqnardSdtdPdt}
\left \{
\begin{array}{l}
dS/dt(t) = -k_1(S(t)+P(t)+E_0-S_0)S+k_{-1}(S_0-S(t)-P(t)) \, \vspace{1.5mm} \\
dP/dt(t) = k_2(S_0-S(t)-P(t)) \,
\end{array}
\right.
\end{equation}
or
\begin{equation}\label{eqnardPdtdEdt}
\left \{
\begin{array}{l}
dP/dt(t) = k_2(E_0-E(t)) \, \vspace{1.5mm} \\
dE/dt(t) =
-k_1E(t)(S_0-P(t)-E_0+E(t))+(k_{-1}+k_2)(E_0-E(t)) \,
\end{array}
\right..
\end{equation}

(\ref{eqnardSdtdEdt}) is often used to analyze the basic model, but
the other two forms are in fact equivalent to it, and sometimes are
more convenient. These systems are nonlinear, and can not be
integrated explicitly. However, they can be further simplified with
the QSSL[\ref{refBanghe2008}].

\smallskip
\textbf{Quasi-Steady-State Law $1$}: Given any small positive number
$\varepsilon>0$, there is a proper positive number $U$ such that
$C(t)$ will go upwards from 0 at $t=0$ to $E_0-\varepsilon$ in a
period less than $\varepsilon$, then it will stay in the interval
between $E_0$ and $E_0-\varepsilon$ until $S(t)/S_0<\varepsilon$, if
$S_0>U$.
\par
\smallskip
\textbf{Quasi-Steady-State Law $2$}: Given any small positive number
$\varepsilon>0$, there is a proper positive number $U$ such that
$|dC/dt(t)|$ will be less than $\varepsilon$ after a fast
initial period less than $\varepsilon$ and keep this state until
$S(t)/S_0<\varepsilon$, if $S_0>U$.
\smallskip

Michaelis-Menten equation is derived from the quadratic equation
$dE/dt=0$, which is assured to be an acceptable approximate solution
of the process after the initial transient period until $S$ is
nearly exhausted provided $S_0\gg E_0$ by QSSLs. This article
provides another equation which approximates the whole process of
the chemical reaction better than $dE/dt=0$ does. This replacement
is first introduced in our former paper [\ref{refBangheincoming}].
In [\ref{refBangheincoming}], we provided an improved method to
measure all rate constants in the simplest enzyme kinetics model
using this replacement with the aid of Michaelis-Menten equation. This method improved the approach in [\ref{refBanghe2009}] greatly.
Here, we do deep analysis of this equation and found that all the
three rate constants in the simplest enzyme kinetics model can be
measured without Michaelis-Menten equation. The results are better
than those gotten from using the Michaelis-Menten equation only,
which shows that this equation can replace the Michaelis-Menten
equation.
\par
The mathematical background can be found in many fundamental books
on mathematical biology[\ref{refFarkas 2001}, \ref{refMurray
20022003}] or ordinary differential equations[\ref{refHirsch and
Smale 1974}, \ref{refHurewicz1958}].
\par
This article is organized as follows. Section 2 introduces the
deviations of Michaelis-Menten curve and Michaelis-Menten equation which is not novel and can be read in many commentaries[\ref{refSchnellMaini2003}].
Section 3 gives our corresponding replacements of the curve and
equation, and the merits for the replacements are given in section
4. Section 5 gives an application of the replacement of the
Michaelis-Menten curve, and the conclusion comes in Section 6. Some
subtle mathematics are left in Appendix.
\section{Michaelis-Menten curve versus Michaelis-Menten equation}

\subsection{Derivation of Michaelis-Menten curve}
Let $t_1$ be the time when the reaction attains its steady-state.
According to QSSL2, after the initial transient, that is $t>t_1$,
the reaction come to the steady-state:
\begin{equation}\label{QSSL2}
dC/dt(t)\approx 0,
\end{equation}
which is equivalent to
\begin{equation}\label{QSSL2dE=0}
dE/dt(t)\approx 0.
\end{equation}
Therefore, during the quasi-steady state of a reaction, the
relationship about the concentrations $S(t)$ and $E(t)$ can be
approximated by the following equation
\begin{equation}\label{QSSL2equal0}
0 = -k_1S(t)E(t)+(k_{-1}+k_2)(E_0-E(t)),
\end{equation}
which yields
\begin{equation}\label{QSSL2equal0transform}
E_0-E(t)=E_0S(t)/(S(t)+K_M).
\end{equation}
\par
We name the curve of enzyme and substrate determined by the equation
(\ref{QSSL2equal0}) or (\ref{QSSL2equal0transform}) as
Michaelis-Menten curve.

\subsection{Derivation of Michaelis-Menten equation}
According to equation (\ref{eqndp}) and the Michaelis-Menten curve
(\ref{QSSL2equal0transform}), we have
\begin{equation}\label{fulltimeMMequation}
dP/dt(t)=k_{2}E_{0}S(t)/(K_{M}+S(t)),
\end{equation}
or equivalently
\begin{equation}\label{fulltimeMMequation2}
v(t)=k_{2}E_{0}S(t)/(K_{M}+S(t)).
\end{equation}
Let $v_0$ denote the initial velocity of the reaction, which is
indeed the velocity when the reaction attains its steady-state, i.
e. $dP/dt(t_1)$. Equation (\ref{fulltimeMMequation}) becomes
\begin{equation}
v_0=V_{\max}S(t_1)/(K_{M}+S(t_1)),
\end{equation}
where $V_{\max}=k_{2}E_{0}$. It may be assumed that
\begin{equation}\label{initialcondition}
S(t)\approx S_0,
\end{equation}
when $0<t\leq t_1$[\ref{refSegelSlemrod1989}, \ref{refSegel1988}]
(A rigorous proof is given in Appendix). Therefore, the
Michaelis-Menten equation is obtained
\begin{equation}\label{MichaelisMentenEquation2}
    v_0=V_{\max}S_0/(K_{M}+S_0).
\end{equation}
Notice that, if $v_0$ is considered as a function of $S_0$, $v_0$ is
increasing and $$\lim_{S_0\rightarrow+\infty}v_0=V_{\max}.$$ This is
why biologists define $k_2E_0$ as $V_{\max}$. They consider it as
the maximal initial velocity. However, as we have shown, it can not
be attained.
\subsection{The determinant of Michaelis-Menten curve}
By distinguishing Michaelis-Menten curve from Michaelis-Menten
equation, we see clearly that the good performance of
Michaelis-Menten equation approximating the real reactions is due to
Michaelis-Menten curve.
\par
Hence, if we find another curve which is a better approximation,
then we can improve the classical Michaelis-Menten equation.
Fortunately, we find one.
\par
The following section gives our better replacements of
Michaelis-Menten curve and Michaelis-Menten equation, respectively.

\section{Replacements of Michaelis-Menten curve and Michaelis-Menten equation}
For brevity here, we just give the formulas of the replacements of
Michaelis-Menten curve and Michaelis-Menten equation, respectively.
Their merits and motivations are given later.
\subsection{Replacements of Michaelis-Menten curve}
The replacement of Michaelis-Menten curve is

\begin{equation}\label{replacementofMM}
E(E_0-E)[k_1SE-k_{-1}(E_0-E)]+SE_0[k_1SE-(k_{-1}+k_2)(E_0-E)]=0.
\end{equation}
We simply denote the left hand side of the above quartic equation as
$A(S,\ E)$.
\subsection{Replacements of Michaelis-Menten equation}
Just like equation (\ref{QSSL2equal0transform}) represents an
explicit solution $E(S)$ of equation (\ref{QSSL2equal0}), there is
an explicit solution of equation (\ref{replacementofMM}) or
$A(S,E)=0$, too. Due to the complexity of the form, we denote
$E(S)=x_2(S)$ here and give its detail in Appendix 7.3.
\par
Hence, we get a replacement for Michaelis-Menten equation.
\begin{equation}\label{eqn:replaceMMeq}
v_0=k_2(E_0-x_2(S_0)),
\end{equation}
where
$x_2=-(36abc-108a^2d-8b^3+12\sqrt{3}(4ac^3-b^2c^2-18abcd+27a^2d^2+4b^3d)^{1/2}a)^{1/3}/12a+(3ac-b^2)/(3a(36abc-108a^2d-8b^3+12\sqrt{3}(4ac^3-b^2c^2-18abcd+27a^2d^2+4b^3d)^{1/2}a)^{1/3})-b/(3a)+(1/2)\sqrt{3}i((36abc-108a^2d-8b^3+12\sqrt{3}(4ac^3-b^2c^2-18abcd+27a^2d^2+4b^3d)^{1/2}a)^{1/3}/6a-2(3ac-b^2)/(3a(36abc-108a^2d-8b^312\sqrt{3}(4ac^3-b^2c^2-18abcd+27a^2d^2+4b^3d)^{1/2}a)^{1/3}))$,
$a=-k_1S-k_{-1}$, $b=(k_1S+2k_{-1})E_0$,
$c=-k_{-1}E_0^2+(k_1S+k_{-1}+k_2)E_0S$ and $d=-(k_{-1}+k_2)E_0^2S$.
\par
The detail form of equation (\ref{eqn:replaceMMeq}) is somewhat
complicated. However, for the purpose of applications, using the
curve (\ref{replacementofMM}) instead of equation
(\ref{eqn:replaceMMeq}) is sufficient.
\par
The following section will show that curve $A(S,E)=0$ approximates
real reactions better than Michaelis-Menten curve does, and hence
the replacement of Michaelis-Menten equation is better than
Michaelis-Menten equation.

\section{Motivation and Derivation of the Replacement}
The replacement of Michaelis-Menten curve was given first in [\ref{refBangheincoming}]. In Section 4 of [\ref{refBangheincoming}], we have given the motivation and derivation of the equation. For the convenience of the readers, we recap those here. For more information, please read [\ref{refBangheincoming}].

This paper adopts the same notations.
To be precisely, they are listed below again.

The first quadrant of the phase plane $S-E$ is divided into five
regions as

\begin{eqnarray*}
  L_1 &=& \{(S,\ E):\ Q(S,\ E)=0,\ S \geq 0\},\\
  L_2 &=& \{(S,\ E):\ P(S,\ E)=0,\ S \geq 0\}, \\
  R_1 &=& \{(S,\ E):\ E > \tilde{E},\ (S,\ \tilde{E})\in L_1\}, \\
  R_2 &=& \{(S,\ E):\ \tilde{E} > E > \hat{E},\ (S,\ \tilde{E})\in L_1,\ (S,\ \hat{E}) \in L_2
  \}, \\
  R_3 &=& \{(S,\ E):\ E < \hat{E},\ (S,\ \hat{E}) \in L_2\}.
\end{eqnarray*}
where
\begin{eqnarray}\label{eqn:P}
  P(S,\ E) &=& -k_1SE +k_{-1}(E_{0}-E), \\ \label{eqn:Q}
  Q(S,\ E) &=& -k_1SE+(k_{-1}+k_2)(E_{0}-E).
\end{eqnarray}

The whole process of the reaction $(S(t),\ E(t))$ can
be drawn on the $S-E$ plane. Since $S(t)$ decreases when $t$
increases, we can consider $E$ to be a function of $S$.
\begin{equation}\label{equationdEdS}
dE/dS=(-k_1SE(S)+(k_{-1}+k_2)(E_0-E(S)))/(-k_1SE(S)+k_{-1}(E_0-E(S)))
\end{equation}

The solutions with its initial condition on the curve $L_2$ will vertically enter the region $R_2$. Then, the concentration of substrate decreases and that of enzyme increases. In fact, these solutions will stay in $R_2$ forever and finally approaches the singular point. For sufficiently large initial concentration of substrate, these solutions go almost horizontally in $R_2$, but at last they will approach the singular point with a certain slope. Therefore, there is an inflection point on each of these solutions.

We have
\begin{equation}\label{twicediffdRdS}
d^2E/dS^2=k_1k_2A(S,\
E)/(k_1SE(S)-k_{-1}(E_0-E(S)))^3.
\end{equation}

Thus, the collection of inflection points satisfies $d^2E/dS^2=0$, that is $A(S,\ E)=0$. As this system satisfies the existence and uniqueness
condition of differential systems, any two different solutions will not intersect. Thus, the curve $A(S,\ E)=0$ is just beneath the real process on the $S-E$ phase plane.

This is how we find the replacement $A(S,\ E)=0$.

\section{Reasons for the replacement being much better}

We will give reasons that (\ref{replacementofMM}) is a much better
replacement of Michaelis-Menten curve in this section.

In [\ref{refBangheincoming}], we have observed that there is a part
of $A(S,\ E)=0$ lying in the region $R_2$ which approximates the
real process well. It is denoted as $L_3$. In fact $L_3$ is the
replacement of Michaelis-Menten curve. Next, we will show that $L_3$
is a better approximation of real reaction than $L_1$ which is the
Michaelis-Menten curve $Q(S,E)=0$ (\ref{eqn:Q}). We only need to
show that $L_3$ approximates real reaction better than $L_1$ does.

\subsection{Comparison in the major process of a reaction}
When the reaction begins, $S$ and $E$ would decrease until they pass
through the curve $L_1$. In this period of the reaction, neither
Michaelis-Menten equation nor its replacement $L_3$ can approximate
the solution well.
\par
Here, the major process of a reaction means that $S$ and $E$ are in
the region of $R_{2}$ excluding the end of the reaction. The
following two subsections will show that $A(S,E)=0$ gives a more
approximation of the real reaction processes than $L_{1}$
(Michaelis-Menten curve) by numerical instances under different
conditions.
\subsubsection{A case when QSSL condition violates} 
\begin{figure}[h]
  \begin{center}
  \includegraphics[width=0.6\textwidth]{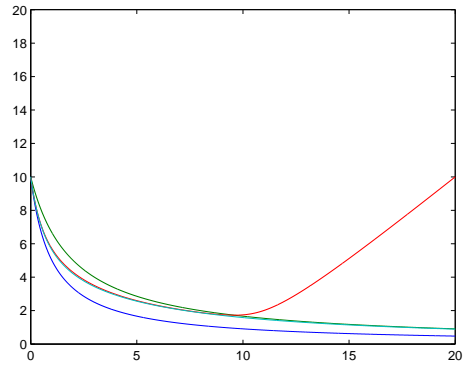}
  \caption{$L_3$ gives a good approximation of the solution after the solution enters the region $R_2$}\label{figuresolutioncompare}
  \end{center}
\end{figure}

we choose
$k_1=1$, $k_2=1$, $k_{-1}=1$, $E_0=10$ but $S_0=20$. In this example
the QSSL can not be used for $S_0$ is not sufficient large compared
with $E_0$. So $L_1$ may not be a good approximation of the
solution. Before the reaction process approaches the region $R_2$,
both approximation of the solution are too bad. However, after the
solution enters the region $R_2$, $L_3$ gives a good approximation
of the solution but $L_1$ doesn't, c. f. Fig
\ref{figuresolutioncompare}.

\subsubsection{Cases when QSSL condition holds} We denote $(S^*(t),\
E^*(t))$ to be the solution with initial condition that $(S^*(0),\
E^*(0))=(S_0,\ E_0)$, $\hat{E}(S^*)$ to be the explicit form of
approximate solution $L_1$ and $\tilde{E}(S^*)$ to be the explicit
form of approximate solution $L_3$. $E^*(t)$ is greater than
$\tilde{E}(S^*(t))$ for all $t>0$, which is proved in Appendix 6.1.
$\hat{E}(S^*(t))$ is smaller than $E^*(t)$ when $t>\hat{t}$, which
is proved by Lemma 3 in [\ref{refBanghe2008}]. Here, $\hat{t}$ is
the time the real process touches the curve $L_3$. That is to say,
after the reaction enters the region $R_2$, the real process lies
between these two approximations. We choose $k_1=1$, $k_2=1$,
$k_{-1}=1$, $E_0=10$ and $S_0=1000$ as the second example. In this
example the QSSL can be used, so $L_1$ is a good approximation of
the solution. During the reaction process, when $S^*(t)<989.8$ or
$t>0.03$, the difference between $\tilde{E}(S^*(t))$ and $E^*(t)$ is
less than the difference between $\hat{E}(S^*(t))$ and $E^*(t)$.
That is to say $L_3$ is a better approximation of the solution after
the initial transient period, i. e. less than 0.03.

We have also done another 250 numerical experiments. $k_1$, $k_2$
and $k_{-1}$ are chosen from $\{1,\ 3,\ 5,\ 7,\ 9\}$, and $E_0=0.5$,
$S_0=20$ or $40$. In each case, we divide $S(\hat{t})$ into 7 equal
pieces with 6 point, which we denote from small to large as $S_1,\
\cdots,\ S_6$. We calculate the distances of $E^*$ and $\hat{E}$ and
the distances of $E^*$ and $\tilde{E}$. Table 1 shows the rate of
these two numbers at the six points. These show that the curve $L_3$
approximate the solution better, and the smaller $S$ is the better
$L_3$ does.

\subsection{Comparison near the ends of the reactions}
Our new equation $A(S,E)=0$, that is $L_3$, approaches the end of
reactions with a tangent line same to that of the reaction
processes, while Michaelis-Menten curve does not. The following is
the proof.
\par
$L_3$ can be regarded as a graph of a function taking $S$ as
independent variable and $E$ as dependent variable. The explicit
form is given in Appendix 7.3. Rewrite $A(S,\ E)=0$ as $\left(
-k_{{1}}S-k_{{-1}} \right) {(E-E_0)}^{3}+ \left( -2\,k_{{1}}SE_{{0}}
-k_{{-1}}E_{{0}} \right) {(E-E_0)}^{2}+ \left(
SE_{{0}}k_{{-1}}-k_{{1}}{E_{{0
}}}^{2}S+k_{{2}}E_{{0}}S+k_{{1}}{S}^{2}E_{{0}} \right)
(E-E_0)+k_{{1}}{S}^{2 }{E_{{0}}}^{2}=0$. Divide each side of the
equation by $S^2$ and let $S\rightarrow0$. Then,
\begin{equation}
k_{-1}(dE/dS(0))^2+(k_1E_0-(k_2+k_{-1}))dE/dS(0)-k_1E_0=0.
\end{equation}
Solving it, we get
\begin{equation}
dE/dS(0)=-\Big(k_{1}E_{0}-(k_{-1}+k_{2})+\sqrt{(k_{1}E_{0}+k_{-1}+k_{2})^2-4k_{1}k_{2}E_{0}}\Big)\Big/2k_{-1}
\end{equation}
where the other root is dropped for the slope must be negative. This
slope is just the slope of the solution $(S^*(t),\ E^*(t))$ entering
the point $(0,\ E_0)$. Thus, this part of $A(S,\ E)=0$ give well
approximation of $(S^*(t),\ E^*(t))$ even when $S(t)$ is very small.
So, we confirm this part of $A(S,\ E)$ is a better approximation of
a real reaction.

\subsection{Comparison of the behaviors for large $S$}

In [\ref{refBangheincoming}], we saw that $L_3$ almost coincide with
$L_1$ when $S$ is sufficiently large. In fact, this can be proved.
$A(S,\ E)$ equals to
$E(E_0-E)[k_1SE-k_{-1}(E_0-E)]+SE_0[k_1SE-(k_{-1}+k_2)(E_0-E)]$. Let
$(\hat{S},\ \hat{E})$ be the point on $L_1$ and $(\hat{S},\
\tilde{E})$ on $L_2$. Then, $A(\hat{S},\
\hat{E})=k_2\hat{E}(E_0-\hat{E})^2>0$ and $A(\hat{S},\
\tilde{E})=-k_2\hat{S}E_0(E_0-\tilde{E})<0$. Thus, there must be one
point $E^*$ between $\tilde{E}$ and $\hat{E}$ such that $A(\hat{S},\
E^*)=0$. This proved that for each $S>0$ there is a point of curve
$A(S,\ E)=0$ lies in $R_2$. In fact, there is only one. The proof is
given in Appendix 7.3.

Moreover, we can prove that, as $\hat{S}\rightarrow +\infty$,
$(E^*-\hat{E})/(E^*-\tilde{E})\rightarrow 0$. Since
\begin{equation}
0=E^*(E_0-E^*)[k_1\hat{S}E^*-k_{-1}(E_0-E^*)]+\hat{S}E_0[k_1\hat{S}E^*-(k_{-1}+k_2)(E_0-E^*)],\nonumber
\end{equation}
it can be proved that
\begin{equation}\label{fractionform}
(k_1\hat{S}E^*-(k_{-1}+k_2)(E_0-E^*))/(k_1\hat{S}E^*-k_{-1}(E_0-E^*))=-E^*(E_0-E^*)/(\hat{S}E_0).
\end{equation}
For
\begin{equation}\label{oncurveL_1}
k_1\hat{S}\hat{E}-(k_{-1}+k_2)(E_0-\hat{E})=0,
\end{equation}
\begin{equation}\label{oncurveL_2}
k_1\hat{S}\tilde{E}-k_{-1}(E_0-\tilde{E})=0
\end{equation}
and
\begin{equation}
k_1\hat{S}E^*-(k_{-1}+k_2)(E_0-E^*)-(k_1\hat{S}\hat{E}-(k_{-1}+k_2)(E_0-\hat{E}))=(k_1\hat{S}+k_{-1}+k_2)(E^*-\hat{E}),\nonumber
\end{equation}
\begin{equation}
k_1\hat{S}E^*-k_{-1}(E_0-E^*)-(k_1\hat{S}\tilde{E}-k_{-1}(E_0-\tilde{E}))=(k_1\hat{S}+k_{-1})(E^*-\tilde{E}),\nonumber
\end{equation}
(\ref{fractionform}) can be written as
\begin{equation}\label{contractionform}
(k_1\hat{S}+k_{-1}+k_2)(E^*-\hat{E})/(k_1\hat{S}+k_{-1})(E^*-\tilde{E})=-E^*(E_0-E^*)/(\hat{S}E_0).
\end{equation}
Letting $\hat{S}\rightarrow +\infty$ on both side of
(\ref{contractionform}),
\begin{equation}
(E^*-\hat{E})/(E^*-\tilde{E})\rightarrow0\nonumber
\end{equation}
for $E^*\rightarrow 0$ when $\hat{S}\rightarrow +\infty$.

Thus, there is a part of $A(S,\ E)=0$ in region $R_2$ asymptotically
approaching to $L_1$ when $S$ approaches $+\infty$.

According to (\ref{contractionform}),
\begin{equation}\label{orderS^3}
E^*-\hat{E}=-E^*(E_0-E^*)(k_1\hat{S}+k_{-1})(E^*-\tilde{E})/(\hat{S}E_0(k_1\hat{S}+k_{-1}+k_2).
\end{equation}
Note that $E^*\leq \hat{E}$, $E_0-E^*\leq E_0$ and
$E^*-\tilde{E}\leq \hat{E}-\tilde{E}$. Because of
(\ref{oncurveL_1}), $\lim_{\hat{S}\rightarrow
+\infty}\hat{E}\hat{S}=K_M$. The difference of (\ref{oncurveL_1})
and (\ref{oncurveL_2}) is

\begin{equation}
k_1\hat{S}(\hat{E}-\tilde{E})-(k_{-1}+k_2)(E_0-\hat{E})+k_{-1}(E_0-\tilde{E})=0.\nonumber
\end{equation}
$\hat{S}\rightarrow +\infty$ implies
$\hat{S}(\hat{E}-\tilde{E})\rightarrow k_2E_0/k_1$. These together
with (\ref{orderS^3}) yields $E^*-\hat{E}=O(1/\hat{S}^3)$.


\section{Application}
The above section has shown that the curve $A(S,E)=0$ approximates
the trajectory $(S(t),E(t))$ of the reaction
(\ref{ChemicalReactionScheme }) better than the curve $Q(S,E)=0$
does. In this section, we will show that $A(S,E)=0$ not only gives
more information about the relationships among the three rate
constants but also gives more accurate evaluations of these
constants.
\subsection{$A(S,E)=0$ gives more information about the rate constants}
For convenience, we repeat $A(S,E)$ and $Q(S,E)$ here again as
\begin{eqnarray*}
  A(S,E) &=& E(E_0-E)[k_1SE-k_{-1}(E_0-E)]+SE_0[k_1SE-(k_{-1}+k_2)(E_0-E)], \\
  Q(S,E) &=& -k_1SE+(k_{-1}+k_2)(E_{0}-E).
\end{eqnarray*}
\par
Rearranging the items of the right side of $A(S,E)$ yields that
\begin{eqnarray*}
  A(S,E) &=& k_{1}(-SE^{3}+E_{0}SE^{2}+E_{0}S^2E) \\
         & + & k_{2}(E_{0}SE-E_{0}^2S) \\
   & +  & k_{-1}(-E^3+2E_{0}E^2+E_{0}SE-E_{0}^2E-E_{0}^2S).
\end{eqnarray*}
\par
By comparing equations $A(S,E)$ and $Q(S,E)$, we find that given
some values of $(S,E)$, we can calculate all the three values of
$k_{1}$, $k_{2}$ and $k_{-1}$ up to a common multiplier by equation
$A(S,E)=0$, but we only get two values of $k_{1}$ and $(k_{-1}+k_2)$
up to a common multiplier by equation $Q(S,E)=0$. In other words,
$Q(S,E)$ only contains the information about the Michaelis constant
$K_{M}$ as a whole, but $A(S,E)$ contains the information of
$a=k_{2}/k_{1}$ and $b=k_{-1}/k_{1}$, which also
yield Michaelis constant by $K_{M}=a+b$.
\par
Moreover, by including an additional equation
\begin{equation}\label{eqn:dpdt}
    dP/dt=k_{2}(E_{0}-E).
\end{equation}
This is just (\ref{eqndp}), from which $k_{2}$ can be measured,
$Q(S,E)=0$ only provides the information about $K_{M}$ and $k_{2}$,
while $A(S,E)=0$ provides that of all the three rate constants
$k_{1}$, $k_{2}$ and $k_{-1}$.

\par
Since equations (\ref{eqndp}) and $Q(S,E)=0$ consist of the origin
of Michaelis-Menten equation, the compounding of equations
(\ref{eqndp}) and $A(S,E)=0$ gives more information about the rate
constants than Michaelis-Menten equation does.
\subsection{$A(S,E)=0$ gives more accurate evaluations of the rate constant}
As an example, we design a numerical experiment to show that
compared with $Q(S,E)=0$, $A(S,E)=0$ not only gives more information
about the relationships among the three rate constants but also
gives more accurate evaluations of these constants. In the example,
we set the rate constants as $k_{1}=0.3$, $k_{2}=0.2$ and
$k_{-1}=0.1$, and the initial concentrations of enzyme and substrate
as $E_{0}=0.5$ and $S_{0}=20$. Some points of $(S,E)$ are measured
on the trajectory of the reaction, and then the results are
calculated by $Q(S,E)=0$ and $A(S,E)=0$, respectively. All the
results are listed in Table (\ref{table2}).
\par
In table (\ref{table2}), we only list the concentrations of
substrate, and do not list the corresponding concentrations of
enzyme for brevity. $a:h:b$ means that the concentrations of
substrate are measured from $S=a$ to $S=b$ with a step length $h$.
After measured these values of $S$ and their corresponding values of
$E$, $K_{M}$ can be calculated by $Q(S,E)=0$ and $A(S,E)=0$,
respectively.
\par
For different sets of points $(S,E)$ as chosen in table
(\ref{table2}), $K_{M}^{A}$ is always closer to its exact value $1$
than $K_{M}^{Q}$. It is even the case, when there are only two
points in the set, such as $(S=3, E=0.1234)$ and $(S=19, E=0.0250)$.
\par
Another phenomena observed from this table is that for both
equations $Q(S,E)=0$ and $A(S,E)=0$, $K_{M}$ is more accurate when
the data set is measured closer to the core region of the steady
state of the reaction. Such a phenomena also gives another support
that both $Q(S,E)=0$ and $A(S,E)=0$ approximate the real reaction
well at the quasi-steady state, moreover $A(S,E)=0$ is better than
$Q(S,E)=0$.
\par
If $k_{2}$ is measured by equation (\ref{eqndp}), then $k_{1}$ and
$k_{-1}$ are all known due to equation $A(S,E)$. In this example, we
assume that $k_{2}=0.200$, and hence, the estimated values of
$k_{1}$ and $k_{-1}$ are listed in the table, too.
\par
Now, we have completely shown that compared with $Q(S,E)=0$,
$A(S,E)=0$ not only gives more information about the relationships
among the three rate constants but also gives more accurate
evaluations of these constants. Thus, we claim that $A(S,E)=0$ is a
better replacement of Michaelis-Menten curve, and combined with
(\ref{eqndp}) gives a better replacement of Michaelis-Menten
equation.

\section{Conclusion}
In this article, we propose another curve that can replace the
Michaelis-Menten curve and another equation that can replace the
Michaelis-Menten equation. We used this new curve to estimate all
the rate constants of the basic enzyme kinetics model. Results show
that this replacement does very well. The Michaelis-Menten curve
only gives information about $K_M$. The Michaelis-Menten equation,
which is derived by combining Michaelis-Menten curve and
(\ref{eqndp}), only gives information about $K_M$ and $k_2$. By
contrasting to Michaelis-Menten curve, the replacement curve gives
more information. And then, the replacement equation gives
information about $k_1$, $k_{-1}$ and $k_2$. Numerical experiments
show that these replacements not only give more information about
the relationships among the three rate constants but also give more
accurate evaluations of these constants.

We did not give the mathematical meaning and reasoning that the
replacement curve gives better approximate than Michaelis-Menten
curve during the major process. Instead, we only give some numerical
examples. We hope to do so in future work.

\section{Appendix}
\subsection{$S(t)$ in the initial transient period of a reaction}
To obtain the Michaelis-Menten equation,
(\ref{initialcondition}) is assumed in former literatures. Here, we
prove it under the conditions in QSSLs. That is to say if $S_0$ is
much more larger than $E_0$, $S(t)$ is nearly equal to $S_0$ when
$0<t\leq t_1$. To be more precise and rigorous, we state it as a
lemma below.

\smallskip
\textbf{Lemma}: Given $E_0$ and any small positive number
$\varepsilon>0$, there is a proper positive number $U$ such that
$|dC/dt(t)|$ will be less than $\varepsilon$ after a fast
initial period $t_1$ less than $\varepsilon$, and keep this state
until $S(t)/S_0<\varepsilon$, if $S_0>U$. Moreover, $|S(t)/S_0|\geq
1-\varepsilon$, for $0<t\leq t_1$.

\smallskip
\textbf{Proof}: The first part of the theorem is just the QSSL2.
According to Lemma 3 in [\ref{refBanghe2008}], $dS/dt(t)<0$
for $t>0$. For equation (\ref{eqnds}),
\begin{equation}\label{inequality}
\begin{array}{ccl}
|dS/dt(t)| &=&
|-k_1S(t)E(t)+k_{-1}C(t)|\\
&=&k_1S(t)E(t)-k_{-1}C(t)\\
&\leq & k_1S(t)E(t)\\
&\leq & k_1S_0E_0.
\end{array}
\end{equation}

Because of QSSL2, we could find $U_1$ and $U_2$ satisfies the
following two statements, respectively. Given $E_0$ and any small
positive number $\varepsilon/(k_1E_0)>0$, there is a proper
positive number $U_1$ such that $|dC/dt(t)|$ will be less
than $\varepsilon/k_1E_0>0$ after a fast initial period less
than $\varepsilon/k_1E_0>0$, and keep this state until
$S(t)/S_0<\varepsilon/k_1E_0>0$, if $S_0>U_1$. Given any
small positive number $\varepsilon>0$, there is a proper positive
number $U_2$ such that $|dC/dt(t)|$ will be less than
$\varepsilon$ after a fast initial period less than $\varepsilon$
and keep this state until $S(t)/S_0<\varepsilon$, if $S_0>U_2$.
Choose $U$ such that $U>U_1$ and $U>U_2$. Then, if $S_0>U$, the
first statement of the theorem is proved. Moreover, for
(\ref{inequality})
\begin{equation}
S(t)/S_0\geq
(S_0-\varepsilon k_1S_0E_0/(k_1E_0))/S_0=1-\varepsilon.
\end{equation}
when $0<t\leq t_1$. This completes the proof.$\Box$

\smallskip

Now
\begin{equation*}
v_0=V_{\max}S(t_1)/(K_{M}+S(t_1))==V_{\max}S_0/(K_{M}S_0/S(t_1)+S_0)\approx
V_{\max}S_0/(K_{M}+S_0),
\end{equation*}
because $1\leq S_0/S(t_1)\leq 1/(1-\varepsilon)$, and
$\varepsilon$ can be arbitrarily small.

\subsection{The convexity of $(S^*(t),\ E^*(t))$}
The solution $(S^*(t),\ E^*(t))$ do not have any inflection point at
all, i. e. $(S^*(t),\ E^*(t))$ do not go across $A(S,\ E)=0$ and
$(S^*(t),\ E^*(t))$ lies above $L_3$. Assume $(S(t),\ E(t))$ is a
solution of system (\ref{eqnardSdtdEdt}), and at time $t_2$ it
intersects with $A(S,\ E)=0$ at $(S(t_2),\ E(t_2))$. Consider
$A(S(t),\
E(t))=E(t)(E_0-E(t))[k_1S(t)E(t)-k_{-1}(E_0-E(t))]+S(t)E_0[k_1S(t)E(t)-(k_{-1}+k_2)(E_0-E(t))]$.
Differentiate it with respect to $t$, and note that $dS/dt=P(S,\
E)$, $dE/dt=Q(S,\ E)$:
\begin{equation}\label{dAdt}
dA/dt =
-Q(E_0-E)P+EQP-E(E_0-E)dP/dt-PQE_0-SE_0dQ/dt.
\end{equation}
For simplicity, we write $A$ as $A(S(t),\ E(t))$, $P$ as $P(S(t),\
E(t))$, $Q$ as $Q(S(t),\ E(t))$, $E$ as $E(t)$ and $S$ as $S(t)$.
Simple calculation shows that
\begin{equation}\label{dPdt}
dP/dt=-k_1PE-k_1SQ-k_{-1}Q
\end{equation}
and
\begin{equation}\label{dQdt}
dQ/dt=-k_1PE-k_1SQ-k_{-1}Q-k_2Q.
\end{equation}
As $A(S(t_2),\ E(t_2))=0$,
\begin{equation}\label{t2equation}
SQE_0=-E(E_0-E)P
\end{equation}
at point $t=t_2$. By putting (\ref{dPdt}), (\ref{dQdt}) and
(\ref{t2equation}) in (\ref{dAdt}),
\begin{equation}
dA/dt(t_2)=-2P(S(t_2),\ E(t_2))Q(S(t_2),\
E(t_2))(E_0-E(t_2)).
\end{equation}
For (\ref{reginor2}), $dA/dt(t_2)>0$. Therefore, if the solution
$(S(t),\ E(t))$ of (\ref{eqnardSdtdEdt}) has one point $t_0$ in the
region $R_2$ satisfying $A(S(t_0),\ E(t_0))>0$, then $A(S(t),\
E(t))\geq0$ for $t>t_0$. We have proved that $A(\hat{S},\
\hat{E})=k_2\hat{E}(E_0-\hat{E})^2>0$, where $(\hat{S},\ \hat{E})$
is on $L_1$. Assume at time $t_4$, $(S^*(t),\ E^*(t))$ reached the
curve $L_1$. Thus, $A(S^*(t_4),\ E^*(t_4))>0$. For continuity, there
is a $\varepsilon$, such that for any $t_4+\varepsilon>t_5>t_4$,
$(S(t_5),\ E(t_5))$ is in the region $R_2$ and $A(S^*(t_5),\
E^*(t_5))>0$. Then, $A(S^*(t),\ E^*(t))>0$ for $t\geq t_4$.
Moreover, $A(S^*(t),\ E^*(t))>0$ for $0<t<t_4$, i. e. $(S(t),\
E(t))\in R_1$, can be verified by straight calculation. According to
(\ref{twicediffdRdS}), $(S^*(t),\ E^*(t))$ is convex.

\subsection{The explicit form}
In this subsection, we talk about the explicit form of the curve
$L_3$. $A(S,\ E)=0$ is a three degree equation of $E$. Thus, for
each $S>0$, there are at most three real solutions of $E$. We have
proved that there is at least one solution of $E$ in the region
$R_2$ for any $S>0$. As in Section 3.2, we have proved that
$A(\hat{S},\ \hat{E})=k_2\hat{E}(E_0-\hat{E})^2>0$ and $A(\hat{S},\
\tilde{E})=-k_2SE_0(E_0-E)<0$. Note that,

\begin{equation}
\lim_{E\rightarrow \infty}A(\hat{S},\
E)/E^3=-k_1\hat{S}-k_{-1}
\end{equation}
for any $\hat{S}>0$. Thus, when $E$ is positively sufficiently
large, $A(\hat{S},\ E)<0$. And when $E$ is negatively sufficiently
large, $A(\hat{S},\ E)>0$. For the continuity of $A(S,\ E)$, there
is at least one real solution greater than $\hat{E}$ and there is at
least one real solution less than $\tilde{E}$. We have already found
three solutions of $E$ when $S>0$, so there are exact three
solutions of $E$ when $S>0$. The explicit form of all these three
solutions, denoted by $E=x_1(S)$, $E=x_2(S)$ and $E=x_3(S)$, can be
given in mathematics.

The three solutions of the equation $ax^3+bx^2+cx+d=0$ are

$x_1=(36abc-108a^2d-8b^3+12\sqrt{3}(4ac^3-b^2c^2-18abcd+27a^2d^2+4b^3d)^{1/2}a)^{1/3}/6a-2(3ac-b^2)/(3a(36abc-108a^2d-8b^3+12\sqrt{3}(4ac^3-b^2c^2-18abcd+27a^2d^2+4b^3d)^{1/2}a)^{1/3})-b/(3a)$,

$x_2=-(36abc-108a^2d-8b^3+12\sqrt{3}(4ac^3-b^2c^2-18abcd+27a^2d^2+4b^3d)^{1/2}a)^{1/3}/12a+(3ac-b^2)/(3a(36abc-108a^2d-8b^3+12\sqrt{3}(4ac^3-b^2c^2-18abcd+27a^2d^2+4b^3d)^{1/2}a)^{1/3})-b/(3a)+(1/2)\sqrt{3}i((36abc-108a^2d-8b^3+12\sqrt{3}(4ac^3-b^2c^2-18abcd+27a^2d^2+4b^3d)^{1/2}a)^{1/3}/6a-2(3ac-b^2)/(3a(36abc-108a^2d-8b^312\sqrt{3}(4ac^3-b^2c^2-18abcd+27a^2d^2+4b^3d)^{1/2}a)^{1/3}))$

and

$x_3=-(36abc-108a^2d-8b^3+12\sqrt{3}(4ac^3-b^2c^2-18abcd+27a^2d^2+4b^3d)^{1/2}a)^{1/3}/12a+(3ac-b^2)/(3a(36abc-108a^2d-8b^3+12\sqrt{3}(4ac^3-b^2c^2-18abcd+27a^2d^2+4b^3d)^{1/2}a)^{1/3})-b/(3a)-(1/2)\sqrt{3}i((36abc-108a^2d-8b^3+12\sqrt{3}(4ac^3-b^2c^2-18abcd+27a^2d^2+4b^3d)^{1/2}a)^{1/3}/6a-2(3ac-b^2)/(3a(36abc-108a^2d-8b^312\sqrt{3}(4ac^3-b^2c^2-18abcd+27a^2d^2+4b^3d)^{1/2}a)^{1/3}))$.

In this problem, $a=-k_1S-k_{-1}$, $b=(k_1S+2k_{-1})E_0$,
$c=-k_{-1}E_0^2+(k_1S+k_{-1}+k_2)E_0S$ and $d=-(k_{-1}+k_2)E_0^2S$.

We have proved that all the three solutions are real, so we should
decide which one represent the curve $L_3$.

We choose $k_1=1$, $k_2=1$, $k_{-1}=1$, $E_0=1$ and $S=1$. Then,
$x_1\approx-0.8892$, $x_2\approx0.6446$ and $x_3\approx1.7446$.
Therefore, $x_2$ is the right one in the region $R_2$. Because
$x_1$, $x_2$ and $x_3$ are continuous functions of $k_1$, $k_2$,
$k_{-1}$, $E_0$ and $S$, and $x_1$, $x_2$ and $x_3$ can not coincide
for any $k_1>0$, $k_2>0$, $k_{-1}>0$, $E_0>0$ and $S>0$, we can
conclude that $E=x_2(S)$ is the explicit form of $L_3$, i. e. the
replacement of the Michaelis-Menten equation.

\par
\noindent $\text{\bf \large Acknowledgments}$
\par
This work is partially supported by a National Key Basic Research Project of China (2011CB302400), by National Natural Science Foundation of China (11301518) and by the National Center for Mathematics and Interdisciplinary Sciences, CAS.

\begin{figure}[h]
  \begin{center}
 \includegraphics[width=0.6\textwidth]{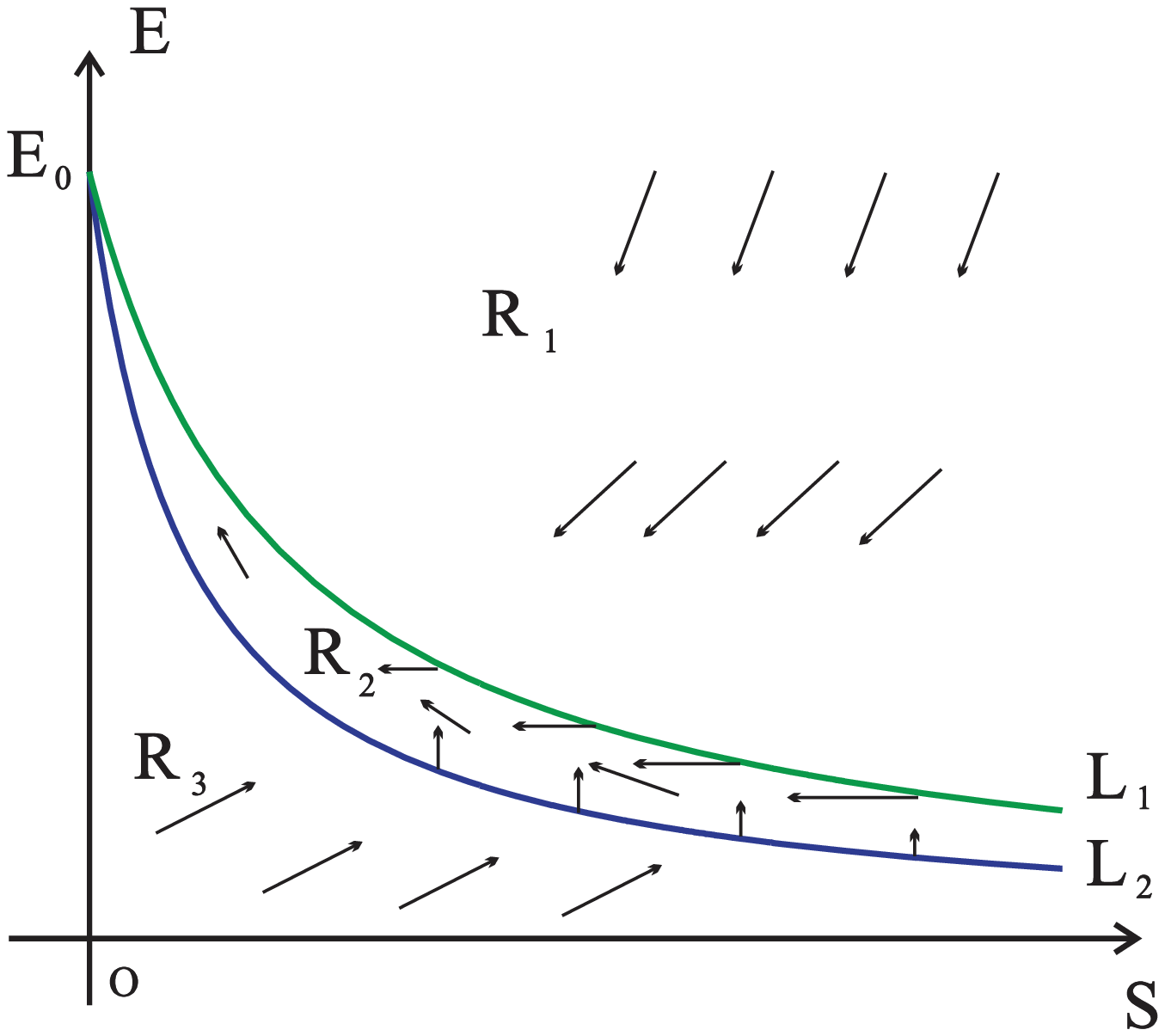}
  \caption{The $S$-$E$ phase plane.}\label{figureSE}
  \end{center}
\end{figure}

\newpage

\newpage
\begin{table}
\centering

\begin{tabular}{l  l l l l l l llll}
  \hline
$k_1$ & $k_2$ & $k_{-1}$ & $S_0$ & $E_0$ & $S_1$ & $S_2$ & $S_3$ & $S_4$ & $S_5$ & $S_6$\\
  \hline
1  & 1 &  1 &  40&  0.5& 1360.849&    978.494&     659.542& 404.013&
211.969&     83.724\\
1  & 1 &  3 &  40&  0.5& 1607.115&    1190.643& 837.895& 549.093&
324.898& 168.645\\
1  & 1 &  5 &  40&  0.5& 1880.447& 1430.567& 1045.033& 724.584&
471.446& 296.762\\
1  & 1 &  7 &40& 0.5& 2182.374&    1700.043& 1283.124& 933.340&
655.858& 476.536\\
1 &1& 9& 40&  0.5& 2514.348& 2000.800& 1554.320& 1178.212& 882.388&
716.419\\
1& 3&   1&   40&  0.5& 532.045& 393.568& 276.394& 180.568& 106.235&
54.160\\
1& 3& 3&   40& 0.5& 623.350& 473.711& 345.536& 239.036& 154.855&
96.309\\
1& 3& 5& 40& 0.5& 724.132& 563.671& 424.992& 308.626& 216.180&
155.887\\
1& 3& 7& 40& 0.5& 834.896& 664.039& 515.482& 390.276& 291.599&
235.622\\
1& 3& 9& 40& 0.5& 956.131& 775.389& 617.712& 484.918& 382.495&
338.216\\
1& 5& 1& 40& 0.5& 372.294& 282.761& 206.082& 142.360& 91.916&
56.441\\
1& 5& 3& 40& 0.5& 432.776& 336.705& 253.673& 183.969& 128.482&
91.841\\
1& 5& 5& 40& 0.5& 499.232& 396.886& 307.883& 232.825& 173.536&
139.448\\
1& 5& 7& 40& 0.5& 571.961& 463.651& 369.141& 289.489& 227.906&
200.886\\
1& 5& 9& 40& 0.5& 651.252& 537.344& 437.867& 354.516& 292.417&
277.761\\
1& 7& 1& 40& 0.5& 308.005& 239.517& 180.320& 130.602& 90.937&
64.393\\
1& 7& 3& 40& 0.5& 355.468& 282.472& 218.981& 165.402& 122.974&
98.218\\
1& 7& 5& 40& 0.5& 407.406& 330.128& 262.678& 205.787& 161.686&
142.000\\
1& 7& 7& 40& 0.5& 464.028& 382.732& 311.712& 252.155& 207.663&
196.886\\

$\cdots$ & $\cdots$ & $\cdots$ & $\cdots$ & $\cdots$ & $\cdots$ & $\cdots$& $\cdots$& $\cdots$& $\cdots$& $\cdots$\\

7&   5&   9&   40&  0.5& 1905.853&    1368.871&    921.202& 562.884&
294.032&     115.236\\
7&   7&   1&   40&  0.5& 1261.696&    893.181& 588.432& 347.446&
170.216& 56.703\\
7&   7&   3&   40&  0.5& 1293.832& 920.402& 610.742& 364.855&
182.747& 64.456\\
7& 7&   5&   40& 0.5& 1326.463& 948.129& 633.572& 382.803& 195.852&
72.877\\
7& 7& 7& 40&  0.5& 1359.591& 976.365& 656.927& 401.299& 209.543&
81.987\\
7& 7& 9& 40& 0.5& 1393.227& 1005.120&    680.817& 420.352& 223.832&
91.812\\
7& 9& 1& 40& 0.5& 1004.972&    714.795& 474.213& 283.225& 141.824&
49.985\\
7& 9& 3& 40& 0.5& 1030.374& 736.374& 491.974& 297.178& 151.996&
56.486\\
7& 9& 5& 40& 0.5& 1056.165& 758.350&     510.144& 311.557& 162.624&
63.526\\
7& 9& 7& 40& 0.5& 1082.346&    780.727& 528.727& 326.368& 173.717&
71.122\\

$\cdots$ & $\cdots$ & $\cdots$ & $\cdots$ & $\cdots$ & $\cdots$ & $\cdots$\\

9&   5&   7&   20&  0.5& 632.410&     458.999&     313.794& 196.824&
108.175&     48.333\\
9&   5&   9&   20&  0.5& 657.240&     480.424& 331.837& 211.529&
119.657& 57.071\\
9&   7&   1&   20&  0.5& 416.007& 297.340& 198.796& 120.367& 62.035&
23.724\\
9& 7& 3&   20&  0.5& 432.973& 311.831&     210.819& 129.935& 69.181&
28.595\\
9& 7& 5& 20& 0.5& 450.325&     326.725& 223.264& 139.952& 76.824&
34.092\\
9& 7& 7& 20& 0.5& 468.080&     342.036& 236.146& 150.433& 84.981&
40.245\\
9& 7& 9& 20& 0.5& 486.247& 357.774& 249.473& 161.389& 93.667&
47.083\\
9& 9& 1& 20& 0.5& 335.698&     241.688& 163.326&     100.608&
53.515& 21.984\\
9& 9& 3& 20& 0.5& 349.216& 253.282& 173.005& 108.384& 59.425&
26.189\\
9& 9& 5& 20& 0.5& 363.046& 265.201& 183.022& 116.521& 65.737&
30.907\\
9& 9& 7& 20& 0.5& 377.191& 277.447& 193.383&     125.026& 72.459&
36.160\\
9& 9& 9& 20& 0.5& 391.664& 290.032& 204.100&     133.911& 79.607&
41.973\\

  \hline
\end{tabular}

\caption{250 numerical experiments.}\label{table1}

\end{table}

\newpage

\begin{table}
  \centering
   \begin{tabular}{cccccc}
  \hline
  S &  $K_{M}^{Q}$ &   $K_{M}^{A}$  &  $k_{1}^{A}$  &   $k_{-1}^{A}$  &
  $k_{2}$\\
  \hline
  3:0.1:19  &   0.996345 &   0.999979  &     0.3052 & 0.1052 &  0.2000 \\
3:0.5:19  &   0.996243 &  0.999978   &     0.3055 &  0.1055 &  0.2000 \\
3:1.0:19  &   0.996112 &  0.999977  &     0.3057 &  0.1057 &  0.2000 \\
3:2.0:19  &   0.995837 &   0.999975 &    0.3063  &  0.1063  &  0.2000\\
3:4.0:19  &   0.995253 &  0.999972 &    0.3073   &  0.1072  & 0.2000 \\
3:8.0:19  &   0.994059 &   0.999972  &     0.3083 &  0.1083  &  0.2000\\
3:16 :19  &   0.991973 &   0.999978 &    0.3088  &  0.1088 &  0.2000\\
         &          &           &          &          &         \\
10:0.1:19 &   0.998506  &  0.999996 & 0.3017 &  0.1017 &  0.2000 \\
10:0.5:19 &   0.998495 &  0.999996 &  0.3017 &  0.1017 &  0.2000\\
10:1.0:19 &   0.998481 &  0.999996 &   0.3018 &  0.1018 & 0.2000\\
10:2.0:19  &  0.998374  & 0.999995  &    0.3019 &  0.1019 &  0.2000\\
10:4.0:19  &  0.998320  & 0.999995  &    0.3019 &  0.1019 &  0.2000 \\
10:8.0:19   & 0.998215  & 0.999996  &     0.3020 &  0.1020 &  0.2000\\
        &        &            &            &             &      \\
16:0.1:19  &  0.999024 &  0.999998  &      0.3011 & 0.1011 &  0.2000\\
16:0.5:19  &  0.999022 &  0.999998  &       0.3011 & 0.1011 &  0.2000\\
16:1.0:19  &  0.999020 &  0.999998  &      0.3011 &  0.1011 &  0.2000\\
16:2.0:19  &  0.998967 &  0.999998  &     0.3011 &  0.1011 &  0.2000\\
  \hline
  \end{tabular}

  \caption{Rate constants estimated by $A(S,E)=0$ or $Q(S,E)=0$. The first
  column indicates the measured concentrations of the substrate during the reaction process,  and the corresponding
  concentrations of enzyme is determined by $S$, so we do not show them explicitly. $a:h:b$ means that
  the concentrations are measured from $S=a$ to $S=b$ with step length $h$. $K_{M}^{Q}$ denotes the Michaelis constant $K_{M}$
  estimated by $Q(S,E)=0$, $K_{M}^{A}$ denotes the $K_{M}$
  estimated by $A(S,E)=0$, $k_{1}^{A}$ and $k_{-1}^{A}$ denote $k_{1}$ and $k_{-1}$ estimated by $A(S,E)=0$, if $k_{2}$
   is provided. Here, we assume that $k_{2}=0.2000$ is exactly estimated by equation (\ref{eqn:dpdt}). }\label{table2}
\end{table}

\end{document}